\documentclass[prb,aps,twocolumn,amsmath,amssymb,floatfix,superscriptaddress]{revtex4}

\usepackage[dvips]{graphics}
\usepackage{color}
\usepackage{soul}
\usepackage[colorlinks=true, citecolor=blue, urlcolor=blue]{hyperref}

\date{\today}

\begin{document}

\title{Tunable flat bands and their signatures in electronic specific heat of an Aharonov-Bohm triangular quantum network}

\author{Sanchita Nandi}

\email{sanchita.nandi$_$r@isical.ac.in}

\affiliation{Physics and Applied Mathematics Unit, Indian Statistical Institute, 203 Barrackpore Trunk Road, Kolkata-700 108, India}

\author{Santanu K. Maiti}

\email{santanu.maiti@isical.ac.in}

\affiliation{Physics and Applied Mathematics Unit, Indian Statistical Institute, 203 Barrackpore Trunk Road, Kolkata-700 108, India}

\begin{abstract}
	
Quantum networks composed of loop-like structures provide a rich platform for exploring electronic transport phenomena and have been 
widely studied in various contexts. However, their thermal response remains relatively unexplored, motivating us to investigate it in 
the present work. We consider a tight-binding (TB) quantum network composed of a finite number of triangular plaquettes, where neighboring 
plaquettes are connected through single bonds and each triangular loop is threaded by an Aharonov-Bohm (AB) flux $\phi$. The interplay 
of nonuniform site coordination, hopping asymmetry, and quantum interference gives rise to both dispersive and completely flat energy 
levels. We derive an analytical condition involving the TB parameters under which one of the energy branches becomes independent 
of the electronic momentum. We further show that the AB flux $\phi$ provides a direct means of tuning the position of the flat band. 
The characteristic features of the energy spectrum associated with this quantum geometry are reflected in the thermal response, which 
we investigate through the electronic specific heat (ESH). In particular, we demonstrate that the position of the flat band manifests 
itself as a distinct feature in the ESH, establishing a direct connection between the flat-band formation and the thermal response of 
the system. Thus, the ESH provides an alternative route for identifying the presence of a flat band, consistent with the analytical 
condition and the corresponding energy dispersion. Moreover, the AB flux allows the ESH to be selectively regulated through the tuning 
of the flat-band position. The proposed approach provides a simple complementary means of identifying momentum-independent energy levels 
and may be extended to other simple and complex quantum networks supporting flat bands.
	
\end{abstract}

\maketitle

\section{Introduction}

Flat band (FB) systems~\cite{model1,model2} have emerged as an important area of research in condensed-matter physics owing to their unusual
electronic properties and the wide range of phenomena associated with highly degenerate energy states. Flat bands~\cite{fb1,fb2,fb3,fb4,fb5} 
can arise through several mechanisms, including lattice geometry~\cite{geo1,geo2,geo3,ekdia1,ekdia2}, sublattice imbalance, and destructive 
quantum interference between different hopping pathways. Their coexistence with momentum-dependent energy levels has attracted considerable 
attention from both theoretical and experimental communities, as it provides a promising setting for exploring diverse quantum phenomena
accompanied by nontrivial topological effects~\cite{topo1,topo2,topo3}. Recent experimental~\cite{exp1,exp2,exp3,exp4} advances have enabled 
the controlled realization of flat bands in a variety of artificial and solid-state systems. Compact localized states have been observed 
in photonic Lieb and rhombic lattices~\cite{lieb1,lieb2,rho}, while optical Lieb lattices~\cite{opt} with ultracold atoms have demonstrated
coherent population transfer into dispersionless bands. Flat-band condensation and disorder-induced localization have also been reported 
in exciton-polariton lattices~\cite{pol}. In addition, artificial electronic lattices constructed by scanning tunnelling
microscopy~\cite{micro1} provide direct control over geometry and hopping pathways. These developments establish engineered lattices 
as versatile platforms for investigating interference-driven flat-band phenomena beyond conventional crystalline materials.

Despite this extensive progress, most investigations of flat-band networks~\cite{fb3,past1,past2,past3,past4,past5}have focused on band
structures, density of states (DOS), compact localized states, disorder-induced localization~\cite{loca1,loca2}, topology, and transport, 
while their thermal behavior has received comparatively little attention. The electronic specific heat, $C_v$, is a well-defined 
thermodynamic quantity~\cite{cv1,cv2,cv3,cv4} that is sensitive to the thermal redistribution of electrons among energy levels within a 
finite energy window around the chemical potential. Consequently, it can depend sensitively on the degeneracy, spacing, and dispersion 
of the single-particle energy spectrum. A transition from dispersive to flat energy levels may therefore leave characteristic signatures 
in $C_v$. To explore this possibility, we consider a finite triangular quantum network~\cite{tri1,tri2} within a tight-binding (TB)
description~\cite{tb1,tb2,tb3}. Adjacent triangular plaquettes are connected through single bonds, while each loop encloses a magnetic 
flux. The interplay among asymmetric nearest-neighbor hopping (NNH) amplitudes within the triangles, coupling between neighboring plaquettes,
and flux-dependent quantum interference~\cite{phi1,phi2,phi3,phi4} allows dispersive and completely flat energy bands to coexist for 
specific hopping parameters and magnetic flux.

In the first part of our work, we present a detailed spectral analysis of the triangular network. We derive an analytical expression 
involving the TB parameters associated with the quantum geometry under which one of the energy branches becomes independent of the 
electronic momentum, thereby establishing the condition for the formation of a flat band. The corresponding dispersion diagrams further
corroborate the analytical results, and we explore how the momentum-independent energy levels evolve systematically 
with the applied magnetic flux. We then investigate the thermal characteristics of the triangular network through the electronic specific 
heat $C_v$. In particular, we inspect the dependence of $C_v$ on the hopping strength, chemical potential, temperature, and magnetic flux, 
and correlate its non-monotonic behavior with the emergence and flux-driven evolution of the flat-band states. The lifting and subsequent
restoration of the flat-band degeneracy give rise to a characteristic peak-dip-peak (PDP) pattern in $C_v$, whose location agrees well 
with the analytically determined flat-band position. Thus, the $C_v$ response provides a direct thermal signature of the flat band and 
offers a complementary means of identifying its position. We further show that this characteristic response persists over a large range of 
parameters, although its magnitude and detailed profile depend on the relative position of $\mu$ with respect to the flat and dispersive 
energy levels.

To reveal the applicability of this thermal signature beyond the simplest triangular geometry, we extend our analysis to triangular 
plaquettes with multiple sites along each arm. The persistence of the flat-band-induced PDP profile in these enlarged triangular loops 
indicates that the characteristic thermal response survives with increasing geometric complexity. As the number of sites and possible 
hopping pathways increases, an exact analytical determination of the flat-band condition becomes progressively cumbersome. In such cases, 
the proposed thermal approach provides a complementary route for identifying flat-band states without requiring a closed-form determination 
of all energy levels. This approach may therefore be extended to a broader class of quantum networks possessing dispersionless energy bands,
provided that the associated flat-band features produce identifiable signatures in the electronic specific heat. With recent advances in
nanofabrication technologies, several theoretically proposed flat-band geometries can now be engineered and investigated
experimentally~\cite{fbexp1,fbexp2,fbexp3}. These developments provide potential platforms for examining the thermal signatures predicted 
in our study. In particular, measurements of the electronic specific heat may offer an experimental route for identifying flat-band states, 
and the persistence of the characteristic response over a broad range of model parameters suggests that the proposed mechanism is not 
restricted to a particular parameter set.

The remainder of this work is organized as follows. Section II describes the AB triangular quantum network, its Hamiltonian, and the 
theoretical framework used to calculate the electronic specific heat. Section III presents a detailed discussion of the spectral features 
and numerical results, with emphasis on the dispersion characteristics, the appearance of flat bands, and their signatures in the electronic
specific heat. Section IV summarizes the main findings of our analysis.

\section{AB quantum network and Theoretical Method}

We first describe the AB quantum network and formulate its TB Hamiltonian. We then outline the theoretical framework employed to calculate 
the electronic specific heat of the system. The spectral properties and the corresponding energy dispersion are discussed separately in 
the subsequent section.

\subsection{Network and the Hamiltonian}

\begin{figure}[ht]
{\centering\resizebox*{8.7cm}{2.9cm}{\includegraphics{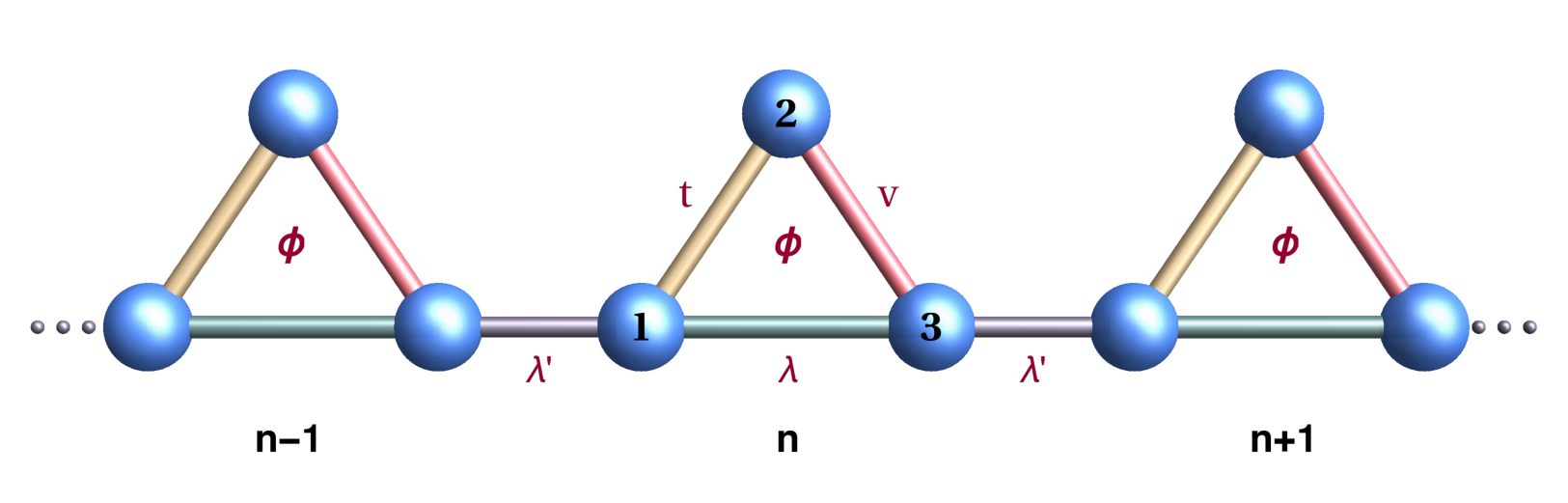}}\par}
\caption{(Color online). Schematic representation of a periodic triangular network with $N$ unit cells. Each triangular plaquette threads 
a magnetic flux $\phi$.}
\label{trifig}
\end{figure}

We consider a tight-binding lattice constructed from a periodic arrangement of interconnected triangular plaquettes, as schematically
shown in Fig.~\ref{trifig}. Each unit cell contains three lattice sites, labeled by $\alpha = 1$, $2$, and $3$ which together form a 
triangular geometry. The Hamiltonian of the system is written as 
\begin{equation}
\begin{aligned}
H &= \sum_{n}\Bigg[\sum_{\alpha=1}^{3}\epsilon\,c_{n,\alpha}^{\dagger}c_{n,\alpha} + 
\sum_{\alpha=1}^{3} \beta e^{i\Theta}c_{n,\alpha}^{\dagger}c_{n,\alpha+1} \\
&\quad + \lambda' c_{n,3}^{\dagger}c_{n+1,1} + \mathrm{h.c.}\Bigg]
\end{aligned}
\label{Ham}
\end{equation}
where $c_{n,\alpha}^\dagger$ $(c_{n,\alpha})$ is the fermionic creation (annihilation) operator at site $\alpha$ of the $n$th triangle.
The index $n$ runs from $1$ to $N$, where $N$ corresponds to the total number of triangular unit cells in the network. The parameter
$\epsilon$ represents the on-site energy, which is assumed to be uniform throughout the lattice. Here, $\beta =t, v, \lambda$ characterize 
the nearest-neighbor hopping amplitudes within a triangular unit cell.
The TB parameter $\lambda'$ represents the hopping strength connecting two adjacent triangles. Each triangular plaquette encloses a magnetic
flux $\phi$, measured in units of the elementary flux-quantum $\phi_0$. Consequently, the hopping amplitudes acquire a phase
factor~\cite{phase1,phase2} $\Theta=(2\pi\phi/3\phi_0)$, where the factor of $3$ in the denominator appears from the three sites in 
each triangular plaquette. The competition among the intra-cell hopping amplitude, inter-cell hopping, and magnetic-flux-induced 
phase modulation give rise to a rich electronic band structure which can be understood form the subsequent discussion.

\subsection{Methodology -- calculation of ESH}

We calculate the electronic specific heat, $C_v$, by differentiating the average energy of the system with respect to temperature,
defined as~\cite{cv1,cv2,cv3,cv4}
\begin{equation}
C_V=\frac{d\bar{U}}{dT},
\label{cvv}
\end{equation}
where the average energy of the system is expressed as
\begin{equation}
\bar{U}=\sum_{i=1}^{3N} (E_i-\mu) f(E_i).
\label{avgeng}
\end{equation}
For a finite triangular network, we find the energy eigenvalues $E_i$ by directly diagonalizing the Hamiltonian matrix of the network. 
The factor $\mu$ in the above expression represents the electrochemical potential and $f(E_i)$ is the Fermi-Dirac distribution function, 
given by $$f(E_i) = \frac{1}{1 + \exp\left[(E_i - \mu)/(k_B T)\right]}$$ where $k_B$ is the Boltzmann constant and $T$ is the absolute
temperature. Substituting $f(E_i)$ into the average energy expression and carrying out the temperature differentiation, we eventually 
obtain the compact expression of ESH as 
\begin{equation}
C_v = \frac{1}{k_B T^2} \sum_i (E_i - \mu)^2 f(E_i)[1 - f(E_i)].
\label{cvfinal}
\end{equation}
The weighting factor $f(E_i)[1-f(E_i)]$ becomes significant only around the vicinity of the electrochemical potential, and thus, the 
dominant contribution to $C_v$ arises from the thermally accessible states within an energy window of order $k_B T$ across $\mu$. 
Hence, we can emphasize that the ESH provides a direct insight into the density of energy states profile.

\section{Results and discussion}

We present and discuss our results in two parts. In the first part, we derive the analytical $E$-$k$ relations and analyze the 
corresponding energy-level diagrams under different input conditions. The second part focuses on the thermal response of the system, 
with particular emphasis on establishing the connection between the characteristic features of the energy levels and the behavior of 
the electronic specific heat.

\subsection{Spectral Analysis} 

We begin by deriving the $E$-$k$ relation for the flux-threaded triangular network. The presence of three lattice sites in each triangular
unit cell leads to a three-component cell basis as 
\begin{equation}
\Psi_n =
\begin{pmatrix}
	c_{n,1} \\
	c_{n,2} \\
	c_{n,3}
\end{pmatrix}.
\end{equation}
In this basis, the TB Hamiltonian can be expressed in a block-matrix form as
{\small\[H=\sum_n\left[\Psi_n^\dagger H_{n,n}\Psi_n+\Psi_n^\dagger H_{n,n+1}\Psi_{n+1}+\Psi_n^\dagger H_{n-1,n}\Psi_{n-1}\right].\]} 
Here, $H_{n,n}$ is equivalent to the effective site energy matrix of the $n$th triangular unit cell, whereas $H_{n,n+1}$ and $H_{n-1,n}$ 
denote the coupling matrices of the $n$th cell to the neighboring ($n+1$)th and ($n-1$)th cells, respectively. The forms of these matrices
are 
\[
H_{n,n}=
\begin{pmatrix}
	\epsilon & t e^{i\Theta} & \lambda e^{-i\Theta} \\
	t e^{-i\Theta} & \epsilon & v e^{i\Theta} \\
	\lambda e^{i\Theta} & v e^{-i\Theta} & \epsilon
\end{pmatrix}
\raisebox{-2.5ex}{,}
\]
\[H_{n,n+1}= H_{n-1,n}^\dagger=
	\begin{pmatrix}
		0 & 0 & 0 \\
		0 & 0 & 0 \\
		\lambda' & 0 & 0
	\end{pmatrix}.
\]
Using the Bloch representation
\[\Psi_n=\frac{1}{\sqrt{N}}\sum_k e^{ikn}\Psi_k\]
we can write
\[H=\sum_k \Psi_k^\dagger H(k)\Psi_k,\]
where
\[
H(k)=H_{n,n}+e^{ik}H_{n,n+1}+e^{-ik}H_{n-1,n}.
\]
The corresponding Bloch Hamiltonian becomes
\begin{equation}
H(k) =
\begin{pmatrix}
	\epsilon 
	& t\, e^{i\Theta} 
	& \lambda\, e^{-i\Theta} + \lambda' e^{-ik} \\[4pt]
	
	t\, e^{-i\Theta} 
	& \epsilon 
	& v\, e^{i\Theta} \\[4pt]
	
	\lambda\, e^{i\Theta} + \lambda' e^{ik} 
	& v\, e^{-i\Theta} 
	& \epsilon
\end{pmatrix}.
\label{hammk}
\end{equation}
The band energies are determined from the equation
\begin{equation}
\det\!\left[ E I - H(k) \right] = 0.
\label{det}
\end{equation}
Using the explicit form of $H(k)$ given in Eq.~\ref{hammk}), Eq.~\ref{det} can be written as the following cubic equation
\begin{equation}
	\begin{aligned}
		(E-\epsilon)^3
		&-(E-\epsilon)A(k)
		-B(k)=0,
	\end{aligned}
	\label{ekrel}
\end{equation}
where
{\small\[A(k)= t^2+v^2+\lambda^2+\lambda'^2 +2\lambda\lambda' \cos\!\left(k-\frac{2\pi\phi}{3\phi_0}\right),\]}
 and
{\small\[B(k)= 2tv\left[ \lambda\cos\!\left(\frac{2\pi\phi}{\phi_0}\right)+\lambda'\cos\!\left(k+\frac{4\pi\phi}{3\phi_0}\right)\right].\]}
 Equation~\ref{ekrel} is the complete analytical $E$-$k$ relation of the flux-threaded triangular network. The three roots of this cubic equation give the three energy bands $E_j(k), j= 1,2,3$. These are given by 
 {\scriptsize
	\begin{equation}
		\begin{aligned}
			E_{1}(k) &= \epsilon
			+ 2\sqrt{\frac{A(k)}{3}}\,
			\cos\!\left[
			\frac{1}{3}\cos^{-1}\!\left(
			\frac{3\sqrt{3}\,B(k)}{2\,[A(k)]^{3/2}}
			\right)
			\right], \\[0.5em]
			E_{2}(k) &= \epsilon
			+ 2\sqrt{\frac{A(k)}{3}}\,
			\cos\!\left[
			\frac{1}{3}\cos^{-1}\!\left(
			\frac{3\sqrt{3}\,B(k)}{2\,[A(k)]^{3/2}}
			\right)
			-\frac{2\pi}{3}
			\right], \\[0.5em]
			E_{3}(k) &= \epsilon
			+ 2\sqrt{\frac{A(k)}{3}}\,
			\cos\!\left[
			\frac{1}{3}\cos^{-1}\!\left(
			\frac{3\sqrt{3}\,B(k)}{2\,[A(k)]^{3/2}}
			\right)
			-\frac{4\pi}{3}
			\right].
		\end{aligned}
	     \label{ek}
	\end{equation}
}
\begin{figure}[ht]
{\centering\resizebox*{6.7cm}{9.4cm}{\includegraphics{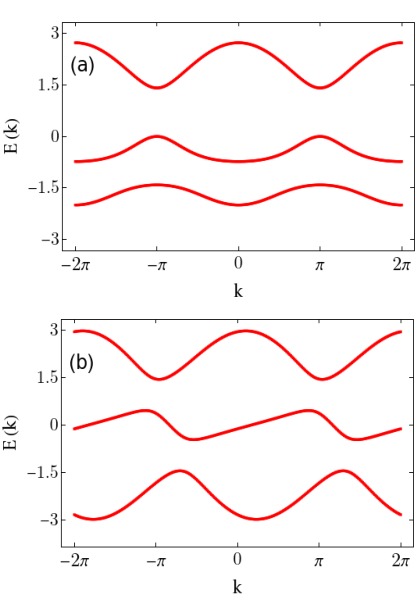}}\par}
\caption{(Color online). Energy band spectra of the triangular lattice for two selected configurations: (a) $\phi=0$, $\lambda=1\,$eV 
and (b) $\phi=0.25$, $\lambda=1.618\,$eV.}
\label{spec2}
\end{figure}
\begin{figure*}[t]
\centering\resizebox{0.7\textwidth}{!}{\includegraphics{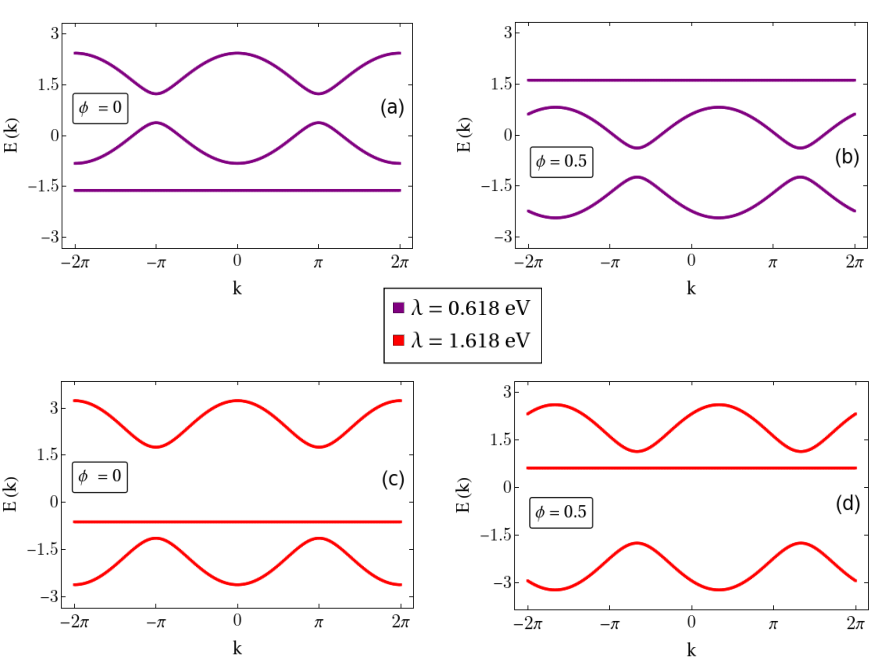}}
\caption{(Color online). Energy band dispersion $E(k)$ at the two flat-band hopping amplitudes, $\lambda=0.618\,$eV and $\lambda=1.618\,$eV, 
for $\phi=0$ and $\phi=0.5$.}
	\label{flatband}
\end{figure*}
\begin{figure}[ht]
	{\centering\resizebox*{8cm}{9.2cm}{\includegraphics{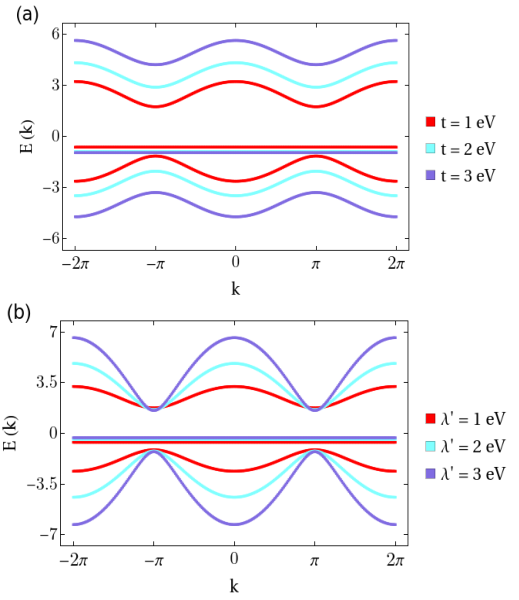}}\par}
	\caption{(Color online). Energy band spectra at $\phi = 0$ for different values of (a) $t$, with $\lambda' = 1\,$eV, and (b) $\lambda'$, with $t = 1\,$eV. In each case, $\lambda$ is set to the respective lower $\lambda_{FB}$ (from Eq.~\ref{fblambda}), while $v$ is maintained at $1\,$eV.}
	\label{speclamt}
\end{figure}
We now study the dispersion characteristics of the flux-threaded triangular lattice by plotting the energy branches $E_j(k)$ over the 
momentum range. Here, the hopping amplitudes $t$, $v$, and  $\lambda'$ are fixed at $1\,$eV, while $\phi$ and $\lambda$ are treated as 
control parameters in order to investigate their role in reshaping the band spectrum. Figures~\ref{spec2}(a) and \ref{spec2}(b) show 
the band structures for two representative parameter choices, namely ($\phi=0, \lambda=1\,$eV) and ($\phi=0.25$, $\lambda=1.618\,$eV), respectively. For $\phi=0$, no magnetic phase is accumulated around the triangular plaquette, and under this condition, the factors
$A(k)$ and $B(k)$, mentioned above, are simplified as
 \[A(k)=t^2+v^2+\lambda^2+\lambda'^2 +2\lambda\lambda'\cos k,\]
and \[ B(k)= 2tv(\lambda +\lambda'\cos k).\] 
In this limit ($\phi=0$), the resulting spectrum becomes perfectly symmetric about $k=0$, reflecting the conventional cosine form of 
the dispersion (Fig.~\ref{spec2}(a)).  
For non-zero magnetic flux, the phase relation between different hopping paths inside the triangular plaquette changes. Consequently, 
the interference between intra-cell and inter-cell hopping processes is altered, resulting in a distinct rearrangement of the energy 
band profile. This effect is clearly seen in Fig.~\ref{spec2}(b). 
The finite dispersion of all three bands indicates that all the energy levels are momentum dependent.

We now focus on the emergence of momentum-independent energy levels, corresponding to the formation of a flat band. Here, we derive an
analytical expression for $\lambda$ in terms of the other hopping parameters, for which the network exhibits a flat band. An exact flat 
band requires the corresponding energy eigenvalues to be independent of $k$. Starting from Eq.~\ref{ekrel}, we derive the flat-band 
condition for $\lambda$, which satisfies the relation
{\small
\begin{equation}
	2 (E - \epsilon)\lambda \lambda' 
	\cos\left( k - \frac{2\pi \phi}{3\phi_0} \right)
	+ 2 t v \lambda' 
	\cos\left( k + \frac{4\pi \phi}{3\phi_0} \right)
	= 0.
	\label{fbcondi}	
\end{equation}}
This relation is fulfilled only at the special flux values $\phi=0$ and $\phi=\phi_0/2$. Consequently, the flat-band formation is 
restricted to these typical flux values. For the zero-flux case, Eq.~\ref{fbcondi} yields the corresponding $k$-independent eigenenergy
\begin{equation}
(E_{FB} - \epsilon)=\frac{-t v}{\lambda}.
\label{fbeng1}	
\end{equation}
$E_{FB}$ refers to the FB energy.
For $\phi=\phi_0/2$, the two phase-shifted cosine terms in  Eq.~\ref{fbcondi} become opposite in sign. Substituting this condition into  Eq.~\ref{fbcondi}, corresponding energy is obtained as
\begin{equation}
	(E_{FB} - \epsilon)=\frac{t v}{\lambda}
	.\label{fbeng2}	
\end{equation}
By evaluating Eq.~\ref{ekrel} at $E = E_{FB}$, we obtain the corresponding flat-band equation
\begin{equation}
\lambda^{4} - \lambda^{2}\left(t^{2} + v^{2} + \lambda'^{2}\right) + t^{2} v^{2}= 0.
\label{finalrel}	
\end{equation}
Solving Eq.~\ref{finalrel}, we get the FB condition which is	
{\small\begin{equation}
\lambda_{FB}^{2} =
\frac{\left(t^{2} + v^{2} + \lambda'^{2}\right)\pm
\sqrt{\left(t^{2} + v^{2} + \lambda'^{2}\right)^{2} - 4 t^{2} v^{2}}}{2}
\label{fblambda}	
\end{equation}}
where $\lambda_{FB}$ refers to the value of $\lambda$ for which the FB emerges. 

For our chosen parameter choice $t=v=\lambda'=1\,$eV, the allowed intra-cell hopping amplitudes are obtained from Eq.~\ref{fblambda} as 
$\lambda_{FB} =0.618\,$eV and $1.618\,$eV. To analyze the flat-band regime, Fig.~\ref{flatband} displays the calculated ($E$-$k$) spectra 
for $\lambda_{FB} =0.618\,$ eV and $1.618\,$eV, under zero-flux and half-flux conditions. The four panels correspond to: (a) $\lambda =0.618\,$eV, $\phi=0$; (b) $\lambda =0.618\,$eV, $\phi=0.5$; (c) $\lambda =1.618\,$eV, $\phi=0$; and (d) $\lambda =1.618\,$eV, $\phi=0.5$. 
In each case, one of the three energy branches becomes completely dispersionless, and it is found that the FB is $N$-fold degenerate.
In the absence of magnetic flux, as shown in panels (a) and (c), the flat-band position follows $E_{FB} = \epsilon-1/\lambda$. Accordingly, 
the FB appears at $-1.618\,$eV and $-0.618\,$eV for $\lambda = 0.618\,$eV and $\lambda = 1.618\,$eV, respectively. In contrast, for 
$\phi = 0.5$, shown in panels (b) and (d), the FB remains intact but shifts to $E_{FB} = \epsilon+1/\lambda$, giving energies of $1.618\,$eV 
and $0.618\,$eV, respectively. This comparison shows that the FB is retained at half-flux-quantum, however, its position is
reversed with respect to $\epsilon$. The dispersionless energy levels arises from compact localization induced by destructive interference
between multiple hopping paths around the triangular loop. At $\phi = 0$ and $\phi = 0.5$, this cancellation condition (Eq.~\ref{fbcondi}) 
is exactly satisfied, leading to momentum-independent energy eigenvalue. However, for intermediate flux values, $0 < \phi < \phi_0/2$,
Eq.~\ref{fbcondi} no longer holds. As a result, the compact localization is weakened and the formerly flat band gains a finite dispersion 
as shown in the Fig.~\ref{spec2}(b).

Following the analysis of $\lambda$-dependent variation of the band spectrum, we next investigate the effects of hopping parameters $t$, 
$v$, and $\lambda'$ at $\phi = 0$. Unless explicitly noted, $v$ is set at $1\,$eV for the entire analysis. Figure ~\ref{speclamt}(a) 
presents the band spectra for $t = 1, 2$, and $3\,$eV, while $\lambda$ is chosen as the corresponding lower $\lambda_{FB}$ obtained from
Eq.~\ref{fblambda} and $\lambda'$ is fixed at $1\,$eV. The red curves, corresponding to $t = 1\,$eV, is the same spectrum shown previously 
in Fig.~\ref{flatband}(c). As $t$ increases, the overall spectral width as well as the energy separation between the sub-bands gradually 
become larger. However, the position of flat band undergoes only a slight shift. A similar spectral evolution is expected when $v$ is 
varied, and therefore the corresponding results are not shown separately.
In comparison, Fig.~\ref{speclamt}(b) shows the band spectra for $\lambda'=1, 2$ and $3\,$eV, with $t$ fixed at $1\,$eV. In each case, 
$\lambda$ is again set to the corresponding lower $\lambda_{FB}$. Unlike the variation with $t$, the parameter $\lambda'$ alters the band
spectrum in a distinct way. Although the total bandwidth increases with $\lambda'$, the energy gaps between the sub-bands remain nearly
unchanged. These findings indicate that $t$, $v$, and $\lambda'$ significantly influence the detailed features of the energy spectrum. 
However, once $\lambda$ is tuned to $\lambda_{\mathrm{FB}}$, one dispersionless energy band persists in the spectrum. This confirms the
robustness of FB emergence against variations in the hopping strengths.

\subsection{Thermal Signatures} 

After analyzing the emergence of exact FBs, we next scrutinize how the resulting band features affect the thermal response of the system. 
We start by defining the units and different factors adopted in the calculation. Temperature is measured in Kelvin (K), and the 
Boltzmann constant is taken as $k_B=8.6173\times10^{-5}\,$eV/K. In obtaining the ESH $C_v$, we impose a fixed total electron number 
$N_{e}$ and determine the corresponding electrochemical potential $\mu$ self-consistently for each parameter set. For the entire ESH
analysis, $t$, $v$ and $\lambda'$ are fixed at $1\,$eV, except where otherwise mentioned, and the results are worked out for $N=300$.

\begin{figure}[ht]
{\centering\resizebox*{6.7cm}{9.5cm}{\includegraphics{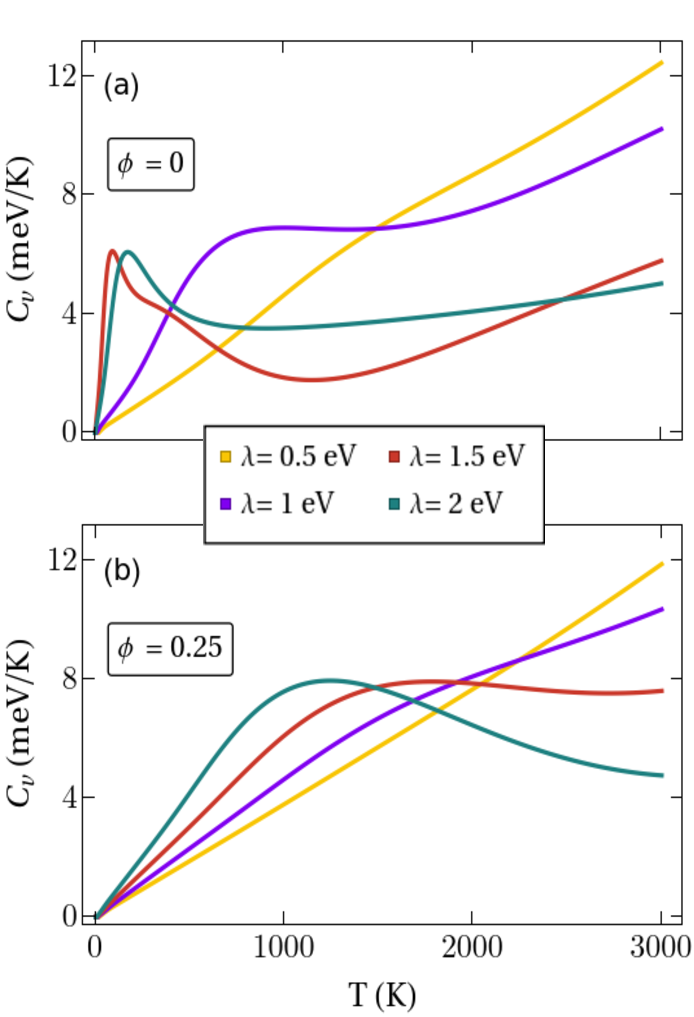}}\par}
\caption{(Color online). Variation of the ESH as a function of temperature for different hopping strengths $\lambda$ under two different
AB fluxes where (a) $\phi = 0$ and (b) $\phi = 0.25$.}
\label{cvT}
\end{figure}

We first focus on Fig.~\ref{cvT}, where $C_v$ is plotted as a function of temperature for selected horizontal hopping strengths in 
the triangular network, comparing the zero-flux case with the finite-flux case ($\phi = 0.25$). Four representative hopping strengths 
are considered those are $\lambda = 0.5, 1, 1.5$, and $2\,$eV. Here, $\mu$ is fixed at the half-filling condition.
In the zero-flux limit, shown in Fig.~\ref{cvT}(a), the influence of hopping strength $\lambda$ is reflected in the different profiles 
of the $C_v(T)$ curves. Since $\mu$ is placed at half-filling, it lies close to the central part of the middle sub-band. 
Therefore, any $\lambda$-induced modification of this sub-band directly affects the number of thermally accessible states near $\mu$. 
As $\lambda$ increases, the middle sub-band becomes increasingly narrow, leading to a stronger concentration of spectral weight near 
$\mu$. Consequently, that gives rise to a larger low-temperature $C_v$ response.
\begin{figure}[ht]
{\centering\resizebox*{6 cm}{14.3cm}{\includegraphics{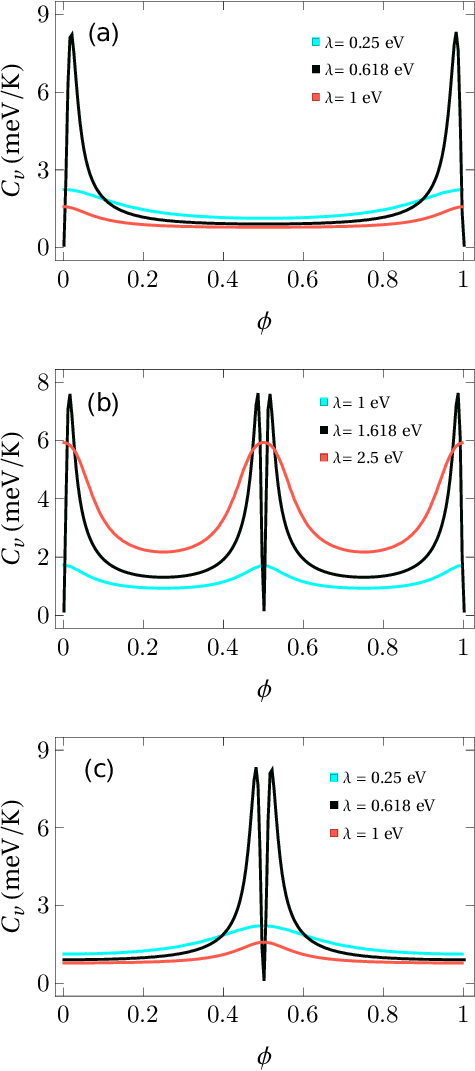}}\par}
\caption{(Color online). Flux dependence of $C_v$ at a fixed temperature $T=200\,$K. Panels (a)-(c) show the response for different chemical-potential positions across the spectrum.}
\label{cvphiall}
\end{figure}
When a finite magnetic flux is introduced, as depicted in Fig.~\ref{cvT}(b), the phase factor modifies the hopping strength and accordingly
the electronic band spectrum. The resulting redistribution of the energy levels around $\mu$ lowers the number of energy states available 
for thermal excitation. As a result, $C_v(T)$ is suppressed at low temperatures relative to the $\phi=0$ case. With increasing temperature, 
the thermally active energy region around $\mu$ widens. Hence, electronic states farther away from $\mu$ also begin to participate in 
the $C_v$ response. In this regime, $C_v$ is no longer primarily governed by the energy levels near $\mu$, but rather by the broader
distribution of energy levels. Consequently, the difference between zero-flux and finite-flux responses becomes less pronounced at 
higher temperatures. It is important to note that the observed behavior strongly depends on the position of electrochemical potential. 
If $\mu$ is shifted away from the half-filling, a different ESH response may emerge, since the reshaping of individual sub-bands with 
varying hopping strength is not uniform across the spectrum.

Figure~\ref{cvphiall}  presents the variation of $C_v$ as a function of the magnetic flux $\phi$ at a fixed temperature, for some
representative horizontal hopping amplitudes $\lambda$ and different positions of $\mu$. As evident from Eq.~\ref{cvfinal}, several 
important features of ESH arise from interplay between the eigen spectrum and the position of $\mu$. We first discuss the quarter-filled 
case, where the chemical potential is chosen within the lower-energy region of the spectrum, as illustrated in Fig.~\ref{cvphiall}(a). 
Under this condition, $C_v$ exhibits pronounced peaks in the vicinity of $\phi \approx 0$ and $\phi \approx 1$. However, exactly at 
these flux values, $C_v$ drops sharply and approaches zero for $\lambda =0.618\,$eV. This behavior follows directly from the flat-band 
spectra in Fig.~\ref{flatband}(a). At $\phi = 0$ and $1$, $\mu$ is aligned with the flat band. Hence, a large number of thermally 
accessible states satisfy $E_i \approx \mu$, causing the factor $(E_i - \mu)^2$ in the $C_v$ expression becomes nearly zero. Although 
the occupation probability $f(E_i)$ remains large, the thermal contribution of these states is strongly suppressed due to the negligible
energy-fluctuation term. This gives rise to clear minima in the specific heat at the flat-band points. When the flux is tuned slightly 
away from these special values, the exact flat-band condition is lifted and the previously degenerate states acquire finite dispersion. 
A large number of states still remain close to $\mu$, but they no longer satisfy $(E_i=\mu)$. Therefore, the factor $(E_i-\mu)^2$ becomes
nonzero while the thermal factor $f(E_i)[1-f(E_i)]$ remains appreciable. The combined effect enhances $C_v$ sharply, producing pronounced 
maxima around $\phi = 0$ and $1$. Thus, the strongest $C_v$ response occurs in the immediate vicinity of the flat-band points rather 
than exactly at them.

However, for $\lambda_{FB} = 1.618\,$eV, a different response is obtained (not shown in the figure). As shown in Fig.~\ref{flatband}(c) and
Fig.~\ref{flatband}(d), the corresponding FB appears near the central part of the spectrum, rather than close to the lower-spectral 
edge. Consequently, at quarter-filling, $\mu$ remains well separated from the FB. Since only states within an energy window of order 
$k_B T$ around $\mu$ contribute significantly to the specific heat, this FB remains thermally inactive and leaves no visible signature 
in $C_v$. For the same reason, the FB response associated with $\lambda_{FB} = 0.618\,$eV at $\phi = 0.5$ (as depicted 
in Fig.~\ref{flatband}(b)), is not reflected in $C_v$ when $\mu$ lies outside the relevant thermal energy window. Therefore, the appearance 
or disappearance of FB signatures in $C_v$ is determined not only by the existence of a FB but also by its proximity to the chemical 
potential. For $\lambda=0.25\,$eV and $\lambda = 1\,$eV, no sharp minima-maxima structure appears in $C_v$. The spectrum remains 
dispersive and the energy levels are spread over a finite range rather than being concentrated at a single energy. However, a comparatively
larger $C_v$ is observed near $\phi = 0$ and $\phi = 1$. This enhancement can be ascribed to the participation of a larger number of 
energy levels in thermal fluctuations, together with a more confined lower band compared with the intermediate-flux cases. Hence, while 
no distinct FB signature emerges for these hopping amplitudes, the overall flux dependence still reflects the increased thermally active
spectral contribution near the integer-flux points.
 
The situation becomes significantly different at half-filling, shown in Fig.~\ref{cvphiall}(b). In this regime, $\mu$ is positioned 
near the center of energy spectrum, allowing the thermally active region to overlap directly with the middle FB. As evident from
Fig.~\ref{flatband}(c) and Fig.~\ref{flatband}(d), $\lambda=1.618\,$eV forms a middle FB at $\phi = 0, 1$ and $0.5$. Accordingly, 
the half-filled case follows the same mechanism discussed earlier. $C_v$ is suppressed at the exact flat-band points and becomes enhanced 
in their vicinity once finite dispersion is restored. Therefore, unlike the quarter-filled case, the half-filled configuration clearly 
exhibits the thermal behavior associated with the $\lambda_{FB} = 1.618\,$eV. Moreover, this signature is observed not only at $\phi = 0$ 
and $1$, but also at $\phi = 0.5$, consistent with the formation of FB at those special flux values. For $\lambda = 1\,$eV and $2.5\,$eV, 
the system lies outside the FB regime. The resulting finite band dispersion broadens the energy-level distribution and weakens the 
accumulation of states within the thermal window of order $k_B T$ around $\mu$. Accordingly, the sharp thermal anomalies characteristic 
of the FB formation are absent for these hopping amplitudes. A similar flux-controlled enhancement is also observed near $\phi = 0$, $0.5$, 
and $1$ for all horizontal hopping strength $\lambda$. Although, the magnitude of this response remains dependent on $\lambda$, since the 
number of thermally accessible states around $\mu$ varies with the hopping strength. In particular, for larger $\lambda$, the middle energy
\begin{figure}[ht]
{\centering\resizebox*{6cm}{14.3cm}{\includegraphics{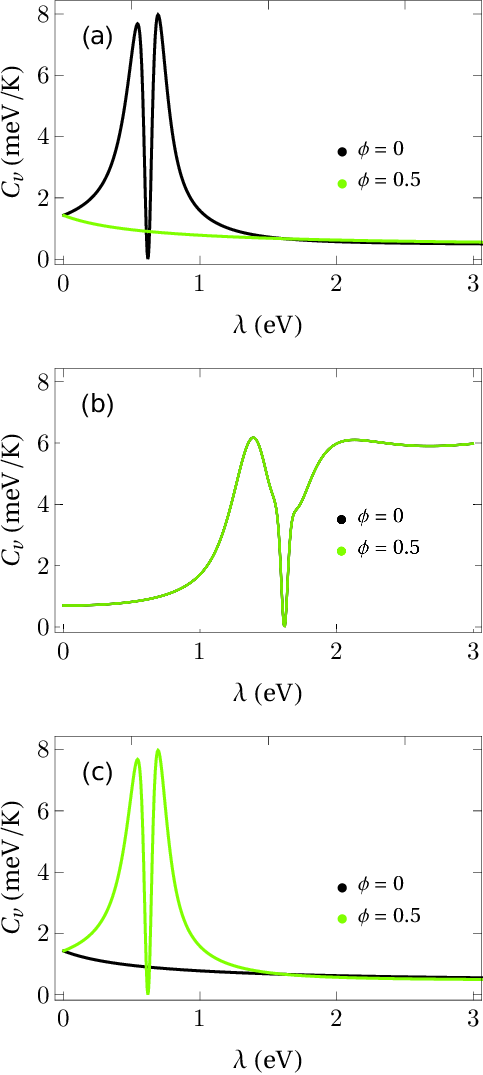}}\par}
\caption{(Color online). Variation of ESH as a function of the hopping strength $\lambda$ at $T = 200\,$K. Panels (a), (b), and (c) 
correspond to different choices of $\mu$, considering both zero and non-zero magnetic flux configurations.}
\label{cvlam}
\end{figure}
bands become more closely aligned within the thermal window, leading to a stronger contribution from energy fluctuations and hence a 
larger $C_v$.

The three-quarter-filled configuration is illustrated in Fig.~\ref{cvphiall}(c). In this regime, $\mu$ shifts toward higher energies, 
and the thermal response is predominantly governed by states located near the upper edge of the spectrum. Therefore, the flat-band-induced
signature in $C_v$ for $\lambda_{FB} = 0.618\,$eV appears only at $\phi= 0.5$, where the flat band lies close to the upper spectral 
boundary, as indicated in Fig.~\ref{flatband}(b). At this flux value, $E_{FB}$ approaches the chemical potential and the same
suppression-enhancement mechanism reappears in $C_v$. Therefore, the thermal response once again reflects the spectral position of the 
flat band relative to $\mu$. By contrast, no comparable feature appears for $\lambda = 0.25\,$eV and $\lambda = 1\,$eV, owing to the 
absence of a corresponding FB and reduced concentration of energy levels within the thermal window around $\mu$. These findings establish 
the ESH as an effective probe for detecting FB formation and determining the corresponding $E_{FB}$ and $\lambda_{FB}$.
 
To further clarify this connection, we next study the ESH as a function of $\lambda$ at the same temperature. This analysis is carried 
out for two representative flux values, $\phi = 0$ and $0.5$, and for three different choices of $\mu$. As shown in Fig.~\ref{cvlam}(a), 
we first consider the quarter-filled configuration, where $\mu$ is located in the lower-energy region of the spectrum. In the absence 
of $\phi$, $C_v$ exhibits an anomalous response at $\lambda = 0.618\,$eV. Based on the preceding band-structure and flux-dependent 
specific-heat analyses, this value is identified as lower $\lambda_{FB}$. A distinct minimum appears between two neighboring maxima, 
indicating that the strongest thermal response occurs in the vicinity of $\lambda_{FB}$ rather than exactly at it. As discussed earlier, 
this anomalous response can be attributed to the reduction of energy fluctuations when the FB level nearly aligns with $\mu$, followed 
by the revival of finite energy fluctuations as the degeneracy is marginally lifted on both sides of $\lambda_{FB} = 0.618\,$eV. Thus, 
the distinctive PDP behavior of $C_v$ provides a clear thermal signature of the FB formation. Beyond this region, the FB-related 
contribution moves away from the thermally active window, and $C_v$ becomes nearly constant. For $\phi = 0.5$, the quarter-filled 
response becomes much weaker. In this case, the upward shift of the FB in the energy spectrum (as shown in Fig.~\ref{flatband}(b)), 
places it far away from the chemical potential $\mu$. Consequently, only a small fraction of states lies within the thermal window of 
order $k_B T$ around $\mu$, resulting in a strongly suppressed $C_v$. Likewise, no anomalous signature is observed at $\lambda = 1.618\,$eV.
This behavior is consistent with the flux-dependent response discussed in Fig.~\ref{cvphiall}(a).
 
The half-filled case, shown in Fig.~\ref{cvlam}(b), exhibits a different dependence of $C_v$ on the hopping strength $\lambda$. 
$C_v$ exhibits a rapid increase with the horizontal hopping strength $\lambda$, followed by a deep valley at $\lambda_{FB}=1.618\,$eV 
and beyond this value, $C_v$ rises again. Notably, the two curves corresponding to $\phi = 0$ and $\phi =0.5$ remain almost converged 
over the entire range. This behavior can be understood from the position of $\mu$. At half-filling, $\mu$ aligns with the FB formed 
at $\lambda_{FB}=1.618\,$eV for both flux setups. Thus, the half-filled configuration uniquely probes the FB response at the same 
horizontal hopping strength in both the absence and presence of $\phi$. Close overlap of the two $C_v$-$\lambda$ curves further suggests 
that the thermal energy window of order $k_B T$ around $\mu$ contains almost the same set of thermally active states for both $\phi = 0$ 
and $\phi = 0.5$. The pronounced suppression of $C_v$ at $\lambda_{FB}=1.618\,$eV therefore reflects the same FB-driven signature in 
both flux configurations as explained in Fig.~\ref{cvphiall}(b).Therefore, the half-filled ESH is primarily sensitive to the middle 
FB at $\lambda_{FB}=1.618\,$eV, while the FB feature at $\lambda_{FB}=0.618\,$eV remains thermally inactive due to its separation from 
the chemical potential.
	 
In the three-quarter-filled case, shown in Fig.~\ref{cvlam}(c), $\mu$ is displaced toward the upper region of the energy spectrum.
\begin{figure}[ht]
{\centering\resizebox*{7.4cm}{12.8cm}{\includegraphics{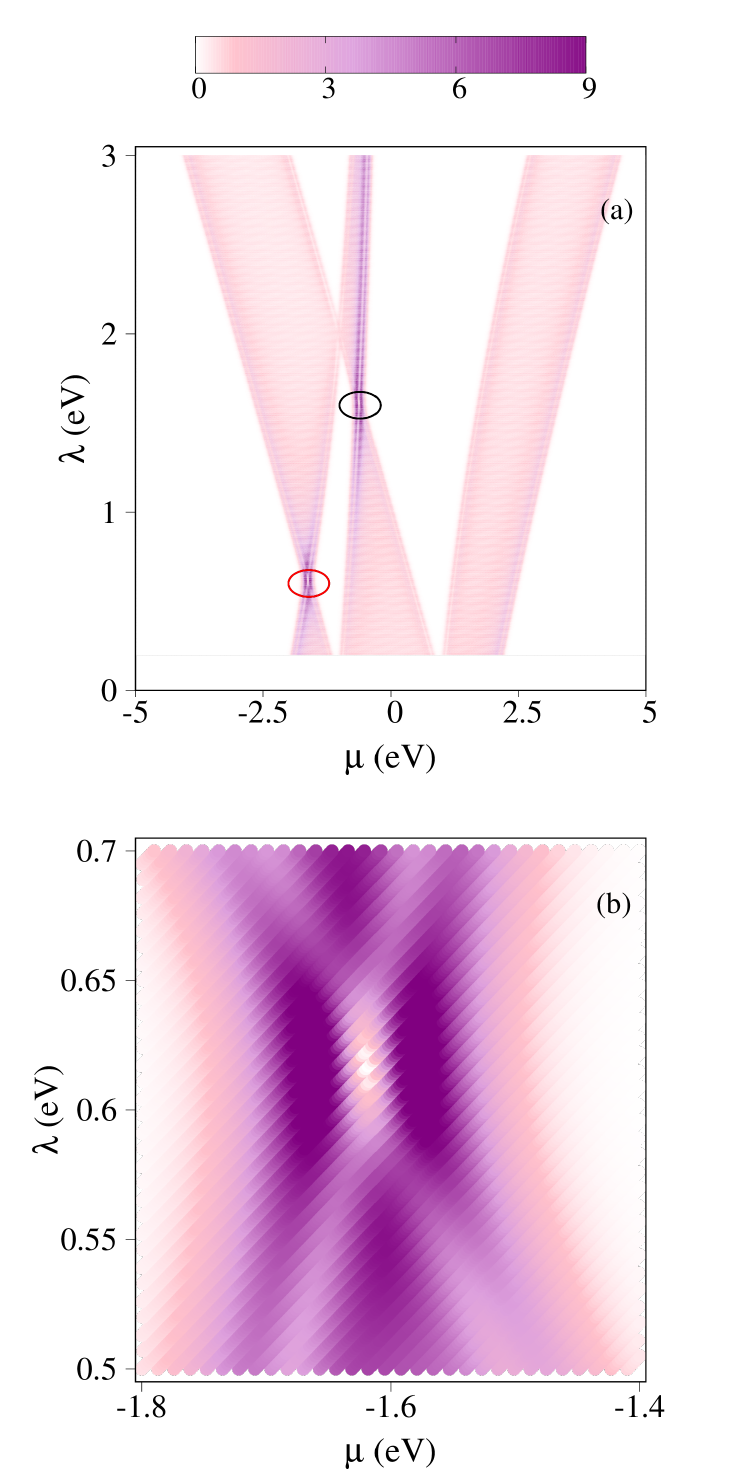}}\par}
\caption{(Color online). (a) Density plot of $C_v$ in the $\mu$-$\lambda$ parameter space at $\phi = 0$. $T$ is set at $200\,$K. 
The color bar defines the magnitude of $C_v$. Panel (b) displays a magnified view of the FB region associated with 
$\lambda_ {FB} = 0.618\,$eV.}
\label{cvmu0}
\end{figure}
As already discussed in Fig.~\ref{cvphiall}(c), the response is reversed relative to the quarter-filled case. For $\phi = 0.5$, the 
FB associated with $\lambda_{FB}=0.618\,$eV lies close to $\mu$, producing a clear FB signature in the $C_v$-$\lambda$ curve. In contrast, 
for $\phi = 0$, the same FB remains far from the three-quarter-filled chemical potential, giving rise to a strongly reduced $C_v$ response. 
This behavior highlights the role of magnetic flux in shifting the FB position relative to $\mu$, thereby controlling its contribution to 
$C_v$. These observations show that the ESH provides a useful route for detecting $\lambda_{FB}$, rather than relying only on explicit 
($E$-$k$) dispersion calculations or analytical FB criteria. The near-vanishing behavior of $C_v$ at $\lambda_{FB}$, together with its 
rapid growth in the adjacent parameter regime, clearly signals the formation of a flat band.

Nevertheless, this detection is selective. For a fixed $\mu$, $C_v$ is sensitive only to FBs that fall within the thermally active 
\begin{figure}[ht] 
\centering \resizebox*{7.5cm}{6.8cm}{\includegraphics{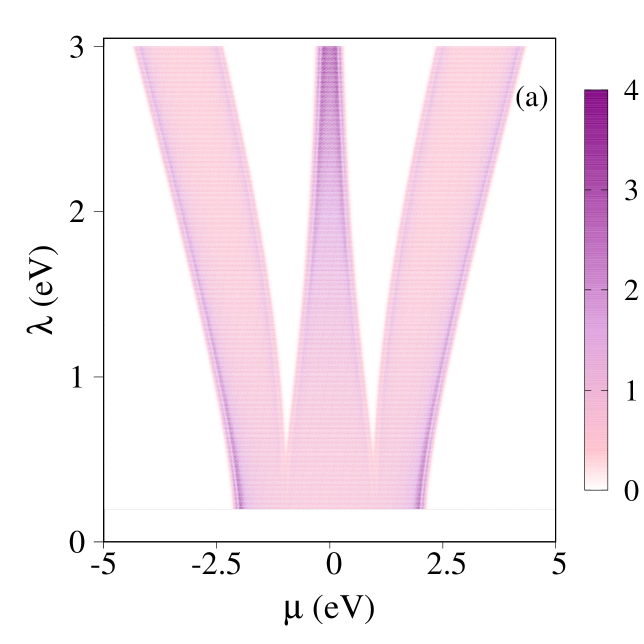}}
\vspace{-0.3cm}
\resizebox*{7.5cm}{6.8cm}{\includegraphics{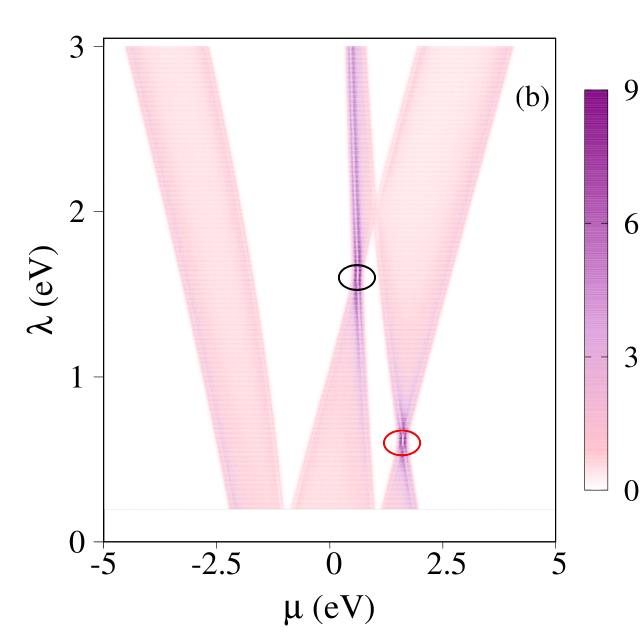}} 
\caption{(Color online). Density plot of $C_v$ as a function of chemical potential and hopping amplitude at $T = 200\,$K for 
(a) $\phi = 0.25$ and (b) $\phi = 0.5$. The FB-induced thermodynamic response is evident in the finite-flux case $\phi = 0.5$.} 
\label{cv0525}
\end{figure}
energy window around $\mu$. The FBs positioned far away from the chosen $\mu$ remain thermally inaccessible and may not appear in 
the $C_v$ profile. For this reason, it is necessary to examine $C_v$ as a function of $\lambda$ while varying $\mu$ across the energy 
spectrum. Such an analysis allows all FB positions to be detected. We investigate $C_v$ as a function of $\mu$ and horizontal hopping 
amplitude $\lambda$ at $\phi = 0$. The resulting density plot is shown in Fig.~\ref{cvmu0}(a). For each point in the $(\mu, \lambda)$ 
plane, the ESH is computed from Eq.~\ref{cvfinal}. The density map exhibits two prominent thermal features near $\lambda_{FB} = 0.618\,$eV 
and $1.618\,$eV, consistent with the band-structure analysis. These regions are marked by the red and black ovals, respectively. Rather 
than appearing as isolated extrema, these points manifest as characteristic thermal patterns consisting of enhanced-$C_v$ regions 
separated by a narrow low-$C v$. A more detailed visualization of the red-oval region is presented in Fig.~\ref{cvmu0}(b). The enlarged 
plot reveals a well-defined PDP profile centered around $\lambda_{FB} = 0.618\,$eV. At the exact FB point, $C_v$ approaches nearly zero 
because the majority of thermally active states satisfy $E_i \simeq \mu$, causing the factor $(E_i-\mu)^2$ to vanish. When $\lambda$ is 
tuned slightly away from $\lambda_{FB}$, the exact degeneracy is lifted and the flat band becomes weakly dispersive. As a result, many
eigenstates still remain within the thermal window determined by the Fermi factor while simultaneously possessing finite energy deviations 
from $\mu$. This collective impact significantly enhances the specific heat, generating distinct maxima on either side of the central 
valley. This unique PDP pattern thus acts as a suitable thermal signature of FB emergence. A similar manifestation is evident in the 
vicinity of $\lambda_{FB} = 1.618\,$eV, though the respective detailed view is excluded for the sake of brevity. The presence of an 
identical thermal profile at both $\lambda_{FB}$ indicates that this behavior is intrinsically rooted in the fundamental generation of 
flat bands, rather than being tied to a specific region of the energy spectrum.

For comparison, we further analyze the density plot of $C_v$ at $\phi = 0.25$ and $\phi = 0.5$, as displayed in Figs.~\ref{cv0525}(a) 
and \ref{cv0525}(b), respectively. Unlike zero-flux case, $C_v$ profile at $\phi = 0.25$ does not show any clear thermal trace of 
FB formation. A similar trend is observed at other intermediate flux, indicating that no FB appears except at $\phi = 0$ and $0.5$. 
In contrast, at half-flux-quantum, the FB feature re-emerges at an energy of equal magnitude but opposite sign to that observed for 
$\phi = 0$, consistent with the corresponding band-structure analysis. The relevant FB regions are indicated by ovals in the density 
plot. However, to keep the presentation concise, the corresponding enlarged view is not shown in the manuscript. A key implication of 
this analysis is that the ESH itself contains sufficient information to accurately locate the FB positions. More importantly, the
characteristic PDP profile enables direct identification of $\lambda_{FB}$ from $C_v$ data alone, without requiring prior knowledge 
of the band structure. 

\begin{figure}[ht]
{\centering\resizebox*{7.5cm}{5cm}{\includegraphics{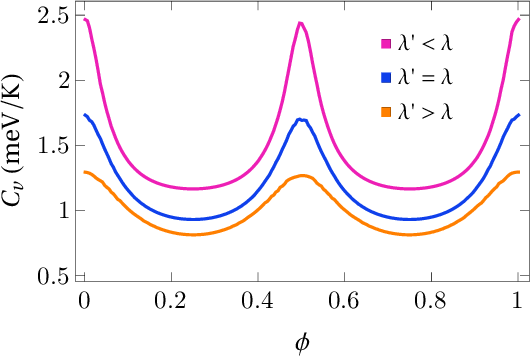}}\par}
\caption{(Color online). $C_v$ as a function of $\phi$ at half-filling for three representative inter-loop coupling regimes: 
$(\lambda' < \lambda)$, $(\lambda' = \lambda)$, and $(\lambda' > \lambda)$.}
\label{cvphilam}
\end{figure}

In addition to the previously discussed $\lambda$-dependent variation of $C_v$, Fig.~\ref{cvphilam} highlights the role of inter-loop
\begin{figure}[ht] 
\centering\resizebox*{7.5cm}{6.9cm}{\includegraphics{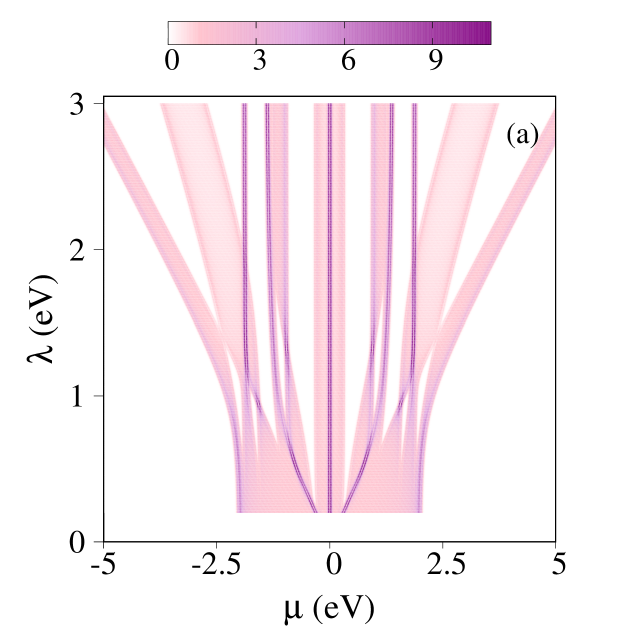}}
\vskip 0.1cm
\resizebox*{7.5cm}{6.9cm}{\includegraphics{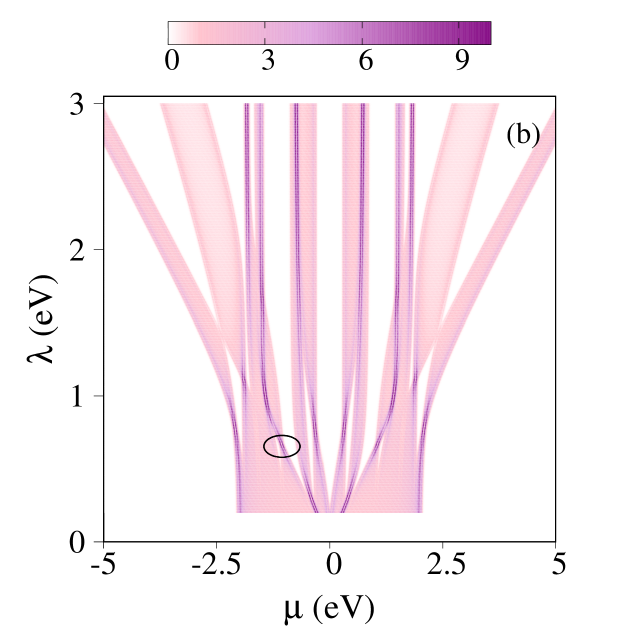}} 
\caption{(Color online). Density map of $C_v$ in the ($\mu$-$\lambda$) plane for a triangular network with five sites per arm at 
$T = 100\,$K, for (a) $\phi = 0$ and (b) $\phi = 0.5$. Multiple FB positions are reflected in the $C_v$ profile.} 
\label{longcv}
\end{figure}
hopping amplitude $\lambda'$, which connects the adjacent triangular plaquettes. Here, we present $C_v$ as a function of $\phi$ for 
different coupling strengths, namely $\lambda'= 0.7, 1.0$, and $1.5\,$eV. For the present calculation, $\lambda$ is fixed at $1\,$eV, 
while $\mu$ is taken at half-filling. Though all three curves preserve the characteristic periodic response with flux, the amplitude 
of $C_v$ is systematically reduced for larger $\lambda'$. An increase in $\lambda'$ strengthens the coupling between neighboring triangular
plaquettes and promotes stronger interaction among their electronic states. As a result, the sub-bands acquire a larger bandwidth. Since 
the total number of electronic states is conserved, redistribution of the same spectral weight over a wider energy range reduces the 
density of energy levels within the thermally active domain around $\mu$. This reduction is responsible for the lower magnitude of $C_v$. 
In contrast, at smaller $\lambda'$, the sub-bands remain comparatively narrow, enabling the density of states to remain more strongly
concentrated near the thermally relevant energy range. This trend is consistent with the behavior shown in Fig.~\ref{speclamt}(b) and 
confirms the role of $\lambda'$ in modifying the specific-heat response.
 
To check the consistency of our results, we now consider a triangular quantum network consisting of $N = 300$ enlarged triangular 
plaquettes, with five sites along each arm. Increasing the size of the triangular geometry provides a more experimentally feasible 
route for attaining a substantial AB flux. For simplicity, we restrict the analysis to five sites along each side of the loop. Compared 
with the earlier triangular geometry, the extended network exhibits a more intricate single-particle spectrum with multiple flat-band 
states. To investigate their thermal signatures, we compute $C_v$ in the ($\mu$, $\lambda$) parameter space at a fixed temperature of 
$100\,$K, taking $t=v=\lambda=1\,$eV. The resulting density maps exhibit several pronounced features associated with the underlying 
FB states. At $\phi=0$, as shown in Fig.~\ref{longcv}(a), one of the FBs remains pinned at zero energy over the entire range of $\lambda$. 
This gives rise to a horizontal ridge of enhanced $C_v$ across the $\mu$-$\lambda$ plane. In contrast, a pronounced modification of the 
spectral structure is observed at $\phi = 0.5$, as presented in Fig.~\ref{longcv}(b). 
\begin{figure}[ht]
{\centering\resizebox*{7.5cm}{5.5cm}{\includegraphics{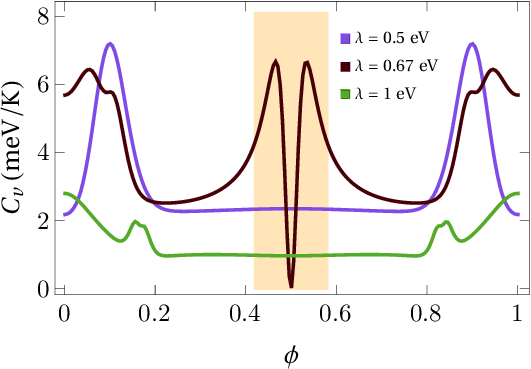}}\par}
\caption{(Color online). Flux dependence of $C_v$ for the enlarged triangular network at $1/3$rd filling and $T=100\,$K. The selected 
values represent the regimes $\lambda < \lambda_{FB}$, $\lambda = \lambda_{FB}$, and $\lambda >\lambda_{FB}$, with 
$\lambda_{FB} = 0.67\,$eV obtained from Fig.~\ref{longcv}. The PDP profile around $\phi = 0.5$ is highlighted by the light orange shading.}
\label{longcvphi}
\end{figure} 
In this case, the FB fixed at $E=0$ is no longer 
present, and the corresponding $C_v$ features are reorganized rather than exhibiting the mirror-symmetric behavior observed for the 
previous network, where each triangle contains three lattice sites. These changes indicate that the applied flux modifies the FB 
positions and, consequently, shifts their thermal signatures across the energy spectrum.

To further validate our observations, Fig.~\ref{longcvphi} shows the variation of $C_v$ with $\phi$ for fixed $\lambda$ at the same 
temperature. We choose $\lambda = 0.5$, $0.67$ and $1\,$eV, where the intermediate value, ($\lambda = 0.67\,$eV), is associated with 
a FB regime marked in Fig.~\ref{longcv}(b). Here, $\mu$ is chosen at one-third filling such that it coincides with the corresponding 
$E_{FB}$. At $\lambda_{FB} = 0.67\,$eV, $C_v$ exhibits a prominent PDP profile around $\phi = 0.5$, which is highlighted by the light 
orange shading. This behavior originates from the flux-driven rearrangement of the energy levels, accompanied by the formation, lifting, 
and subsequent recovery of highly degenerate states. By comparison, when $\lambda$ deviates from $\lambda_{FB}$, as in the cases of 
$\lambda = 0.5\,$eV and $1\,$eV, the corresponding $C_v$ curves do not show an equally pronounced feature. These results demonstrate that 
the ESH remains a sensitive probe of flat-band formation even in enlarged triangular networks containing many sites and multiple energy 
bands. The additional sites generate several FBs, which appear as distinct structures in the $C_v (\mu,\lambda)$ landscape and can be 
shifted through the applied flux. Therefore, $C_v$ offers an effective way to track both the emergence and flux-controlled displacement 
of flat bands.

\section{Closing Remarks}

Our present study analyzes the spectral and thermal characteristics of a flux-threaded triangular network within the tight-binding
approximation. We first investigate the energy spectrum to elucidate its dependence on the applied magnetic flux and the different 
hopping strengths. We obtain an analytical criterion for the intra-cell hopping amplitude at which a flat band emerges and subsequently 
inspect the evolution of the corresponding flat-band states as the magnetic flux through the triangular plaquettes is varied.
We next investigate the thermal behavior of the system by evaluating the electronic specific heat. Our results show that the magnetic 
flux modifies the thermal response in a nontrivial manner, particularly near specific values of the intra-triangle hopping amplitude. 
A detailed analysis of the hopping-amplitude dependence of the electronic specific heat under zero-flux and half-flux quantum conditions
provides further insight into this behavior. The electronic specific heat, $C_v$, exhibits nearly vanishing minima at specific 
intra-plaquette hopping strengths, whose values closely agree with those obtained from the analytical results. In addition to the 
hopping amplitude and applied flux, the choice of chemical potential plays a decisive role in determining the thermal response. 
A flat band produces a visible signature in $C_v$ only when its energy lies within the thermally relevant energy window around $\mu$, 
otherwise, its contribution to the specific-heat profile remains negligible. We therefore investigate the combined effects of the 
chemical potential and intra-triangle hopping amplitude on $C_v$, which establishes a more complete connection between the energy 
spectrum and the thermal response. The corresponding density map exhibits a distinctive thermal pattern characterized by a low-$C_v$ 
valley enclosed by two enhanced-$C_v$ regions. The positions of these features remain consistent with those of the flat bands obtained 
from the spectral analysis. This characteristic PDP behavior originates from the suppression of energy fluctuations as the flat-band 
energy approaches the chemical potential. Slight deviations from the flat-band position lift the exact degeneracy and restore finite 
energy fluctuations, thereby producing specific-heat peaks on either side of the central minimum.
 
The persistence of the characteristic PDP structure in larger triangular geometries indicates that the thermal signature also survives 
with increasing system size, thereby supporting the applicability of the proposed approach beyond the simplest triangular network.
By monitoring $C_v$ as a function of magnetic flux, chemical potential, and hopping strength, the evolution of flat-band contributions 
can be followed without requiring a complete determination of the electronic band structure. More importantly, the present results 
demonstrate that the electronic specific heat can serve as a sensitive thermal probe of flux-driven shifts, restructuring, and loss of 
flat-band states, provided that the corresponding flat-band energies lie within the thermally relevant energy window around $\mu$. Beyond
triangular networks, the present findings may also be relevant to other quantum systems supporting flat bands, indicating that the 
electronic specific heat provides a complementary approach for investigating flat-band behavior.

\end{document}